\documentclass{article}

\usepackage{arxiv}

\usepackage[utf8]{inputenc} 
\usepackage[T1]{fontenc}    
\usepackage{hyperref}       
\usepackage{url}            
\usepackage{booktabs}       
\usepackage{amsfonts}       
\usepackage{nicefrac}       
\usepackage{microtype}      
\usepackage{lipsum}
\usepackage{graphicx}
\usepackage{natbib}
\usepackage{float}
\usepackage{amsmath,amssymb,array}
\usepackage{longtable}
\graphicspath{ {./images/} }

\title{Using Time Series Measures to Explore Family Planning Survey Data and Model-based Estimates
\thanks{\textit{\underline{Citation}}: 
\textbf{Authors. Title. Pages.... DOI:000000/11111.}} 
}

\author{
  Oluwayomi Akinfenwa\\
  Hamilton Institute\\
  Maynooth University\\
  Maynooth\\
  Co. Kildare, Ireland \\ 
  \texttt{oluwayomiakinfenwa@gmail.com} \\
   \And
   Niamh Cahill\\
  Department of Mathematics and Statistics\\
  Maynooth University\\
  Maynooth\\
  Co. Kildare, Ireland \\
  \texttt{Niamh.Cahill@mu.ie} \\
  \And
 Catherine Hurley\\
  Department of Mathematics and Statistics\\
  Maynooth University\\
  Maynooth\\
  Co. Kildare, Ireland \\
  \texttt{Catherine.Hurley@mu.ie} \\
}

\begin{document}

\maketitle

\begin{abstract}
Family planning is a cornerstone of reproductive health and a global development priority, central to Sustainable Development Goal indicator 3.7.1 on meeting women’s need for modern contraception. Monitoring progress toward universal access to family planning is challenged by substantial temporal gaps in household survey data across countries. To address this, the United Nations Population Division employs the Family Planning Estimation Model (FPEM), a Bayesian hierarchical time series model that produces annual estimates and projections of key family planning indicators. FPEM leverages the hierarchical structure of countries nested within regions and subregions, allowing information to be shared across levels to improve estimates in countries with sparse data. While FPEM enables consistent monitoring, its estimates are inherently smoothed, making it important to assess how well modelled trends reflect the underlying survey observations and the heterogeneity of family planning trajectories across countries and regions.
 
This paper examines the alignment between FPEM model-based estimates and observed survey data on modern contraceptive use, a key family planning indicator. The methods use selected time series diagnostic indices from the \textit{wdiexplorer} R package, which explicitly accounts for pre-defined grouping structures, in this case countries nested within sub-regions. These indices quantify variation, trend, and temporal structure in country-level panel data while enabling comparison within and across sub-regional groupings. Visualisation is central to the analysis, enabling direct comparison of relatively sparse survey observations with annual modelled trajectories and facilitating the interpretation of diagnostic measures at both country and sub-regional levels. By jointly visualising survey data and model estimates, as well as residual-based diagnostics, the study identifies where modelled family planning trends closely align with observed patterns and where meaningful discrepancies arise. The results highlight the value of group-aware exploratory visualisation for improving transparency, interpretability, and assessment of model-based estimates. All analyses are fully reproducible and publicly available.   
\end{abstract}


\section{Introduction}\label{introduction}
The right of any individual to decide freely and responsibly on the number and spacing of their children has long been recognised as a fundamental human right \citep{bracke2022women} and is central to improving lives and building equitable societies \citep{unitednations2019}. This positions improvements to family planning as a cornerstone of progress as reflected in Sustainable Development Goal indicator 3.7.1: ``Proportion of women who have their need for family planning satisfied with modern methods''. Family planning is therefore a key global development priority, integral to reproductive health services and broader socio-economic development \citep{bracke2022women}. A range of family planning programmes operate at both global and national levels to expand access to reproductive health services, strengthen health systems, and promote informed contraceptive choice. Among these initiatives is Family Planning 2030 (FP2030) \citep{FP2030}, which supports the global commitment to achieving universal access to family planning by 2030.

FP2030 evolved from the FP2020 initiative, launched at the 2012 London Summit on family planning with a vision tagged ``$120 \; \text{by} \; 20$'' numeric vision, which sought to enable 120 million additional women and girls access to modern contraception by 2020 \citep{hardee2021advancing}. As the FP2020 initiative concluded in 2020, it gave way to FP2030 with a broader vision to promote the right of women and girls to lead healthy lives and make informed choices about contraception \citep{FP2030}. This is an initiative that is closely aligned with the global commitment of universal access to sexual and reproductive health under SDG 3.7.

Despite the global commitment to family planning, progress varies widely across countries and regions, partly due to differences in access, program implementation, and socio-demographic factors \citep{kuang2016global}. To monitor family planning progress, data are collected through household surveys such as, the Demographic and Health Survey(DHS), Performance Monitoring for Actions(PMA), UNICEF Multiple Indicator Cluster Surveys (MICS), national surveys, and other surveys \citep{unitednations2019}. These household surveys are typically conducted every 3 to 5 years, resulting in temporal gaps of up to 6 years for certain countries. In some rare cases, national surveys are conducted annually or semi-annually \citep{cahill2021using}.

Temporal gaps in large-scale survey data make consistent monitoring of family planning progress challenging. To address this limitation, the Family Planning Estimation Model (FPEM) was developed \citep{alkema2013national, cahill2018modern}. FPEM produces annual survey-informed estimates and projections for key family planning indicators using Bayesian hierarchical time series inference. Where survey data are available, the model interpolates estimates and it projects estimates for data gaps between infrequent survey observations. FPEM employs Bayesian hierarchical inference to share information across hierarchical structures, countries nested in subregions, subregions in regions, and regions within the global level. The model estimates are smoothed by design, because they tend to reflect the hierarchical overall behaviour, which makes it important to explore how these modelled trends align with the underlying survey observations.

This paper examines the extent to which modelled estimates from FPEM align with the underlying survey data, using selected diagnostic indices from the \textit{wdiexplorer} package \citep{akinfenwa2025wdiexplorerrpackagedesigned}. These indices are time series measures designed to quantify variation, trend, shape and sequential temporal features of country-level panel data. By using \textit{wdiexplorer} to jointly analyse the survey observations and the corresponding modelled trajectories, as well as evaluating the residuals, the analysis identifies where the model aligns with the observed data and where notable differences arise. This facilitates the identification and comparison of temporal behaviours of the data and the associated model trends, enhancing understanding of patterns in modern contraceptive use across countries. All analyses and visualisations presented in this paper are fully reproducible and available on GitHub at \url{https://github.com/Oluwayomi-Olaitan/Family-Planning-Exploratory-Analysis}.

The remaining sections of this paper are organised as follows. Section 2 provides background to the study, outlining the family planning survey data, the model-based estimates, and the \textit{wdiexplorer} methodology. Section 3 describes the data processing steps for the survey dataset and the dataset of estimates and presents an initial visualisation of the survey data. Section 4 introduces the exploratory analysis of the survey data and model-based estimates using selected \textit{wdiexplorer} diagnostic indices. Section 5 presents the residual-based exploratory analysis. Finally, Section 6 concludes with the findings, benefits of assessing how well the model-based trends align with the underlying survey observations, the limitations of the diagnostic indices in the context of family planning data, and directions for future work.

\section{Background}\label{background}
In this section, we discuss the family planning survey data, the model-based estimates, and our \textit{wdiexplorer} methodology. The survey and model-based data were sourced from the United Nations Population Division (UNPD), while \textit{wdiexplorer} offers a workflow with time-series diagnostic measures and visualisation tools for exploring these measures to identify patterns, outliers, and other notable features.

\subsection{Family Planning Survey Data}
According to \citep{hubacher2015definition}, modern contraceptive methods include medical products or procedures such as pills, intrauterine devices, implants, condoms, and sterilisation, which are designed to reliably prevent pregnancy. Traditional contraceptive methods include behavioural and natural approaches, such as withdrawal, rhythm methods, douching, and folk methods. The prevalence of either modern or traditional contraceptive methods is defined as the proportion of the target population, typically women of reproductive age (15 - 49) using that method at a given time relative to the total population of the same group. These proportions (prevalence of modern and traditional contraceptive use) are the primary variables of interest for our analysis.

These proportions alongside other demographic variables are captured through family planning surveys. A collation of family planning survey data is organised and maintained by the UNPD in the World Contraceptive Use database \citep{UNPD_family_planning_data} and updated annually, providing country-level observations of family planning indicators to support the monitoring of family planning progress. The dataset provides survey observations of contraceptive use (modern and traditional) and unmet need, obtained from nationally representative household surveys, including Demographic and Health Surveys (DHS), UNICEF Multiple Indicator Cluster Surveys (MICS), Performance Monitoring for Action (PMA), Reproductive Health Surveys, Contraceptive Prevalence Surveys (CPS), national surveys, and other sources \citep{cahill2018modern}.

The dataset spans 197 countries, including Taiwan and Kosovo, which are not universally recognised as sovereign nations, and covers 1950 to 2021, although coverage and completeness vary across countries, time, and indicators. Survey observations are disaggregated by data series type, capturing the survey source, union status, indicating whether individuals are married/in a union or not, alongside other demographic characteristics and measures of observation uncertainty in the form of standard errors.

\subsection{Family Planning Model-based Estimates}
The Family Planning Estimation Model (FPEM) is a Bayesian hierarchical time series model used by the UNPD to generate annual estimates and projections of key family planning indicators \citep{alkema2013national, cahill2018modern} using survey data as input. The model leverages the Bayesian inference framework, where country-level estimates are informed by subregional, regional, and global trends. This structure allows reliable projections even in countries with sparse data, as information is effectively borrowed from nested hierarchical levels. Its structure consists of a process model that captures long-term changes in contraceptive prevalence through logistic growth curves reflecting the realistic assumption that growth accelerates at intermediate levels of use and slows as prevalence reaches saturation \citep{alkema2013national, cahill2018modern}. 

FPEM serves as the core family planning  estimation framework within the UNPD, providing projections for family planning indicators and is implemented in the family planning estimation tool for country-level use, available in the R package \textit{fpemlocal} \citep{guranich2021fpemlocal}. While FPEM is central, there are other methodologies employed to complement it in assessing program inputs and impact and setting strategic targets.

The model estimates of family planning indicators are available on the UNPD website \citep{UNPD_family_planning_data} and the dataset contains estimates from 1970 to 2030. There are estimates for 14 indicator levels, with both numerical counts and percentages for contraceptive prevalence, demand for family planning, and unmet need, classified by method type (any, modern, traditional). Each record is associated with a \textit{LocationID}, corresponding to the division numeric code and \textit{Location}, country name, with \textit{Time} representing the year. The dataset also includes \textit{Variant} that captures the estimates and their credible intervals, by 95\% lower bound, 80\% lower bound, median estimate, 80\% upper bound, and 95\% upper bound. It contains the \textit{Age range} information, \textit{Categories}, with 3 population groups  (``All women'', ``Married or in a union women'', and ``Unmarried women''), and \textit{Estimate method} which indicates whether the estimate is interpolated for observed data periods or projected for periods without data.

These model-based annual estimates serve as a dataset to monitor progress toward global family planning goals, such as the FP2030 initiative, and other key family planning targets. The dataset provides consistent and comparable estimates across countries, sub-regions, and regions over time.

\subsection{\textit{wdiexplorer:} Time Series Diagnostic Measures} \label{widexplorer}
To assess and explore the family planning survey data related to contraceptive use and corresponding model-based annual estimates generated by FPEM, we apply our \textit{wdiexplorer} R package to these two datasets. \textit{wdiexplorer} \citep{akinfenwa2025wdiexplorerrpackagedesigned} is a time series exploratory tool originally designed for exploratory analysis of country-level panel data of the World Development Indicators (WDI), as the name implies. It can also be applied to any dataset formatted as repeated observations over time and across countries. The package provides a structured workflow for data exploration through diagnostic indices and visualisation tools. By explicitly incorporating pre-defined grouping structures, such as geographical region, income, or other categories, \textit{wdiexplorer} facilitates the identification of patterns, outliers, and variations within and across group levels, as well as temporal and shape-based features.

In \textit{wdiexplorer}, we use a three-stage process designed for the exploratory analysis of country-level panel data. The first stage involves obtaining data from the WDI, or another source, and assessing its completeness and validity. This stage uses visual summaries to assess data availability and missingness across countries and years, grouping countries by a predefined variable to compare patterns within and across groups.
The second stage computes a set of ten diagnostic indices that quantify key characteristics of the data series. Inspired by the scagnostics framework \citep{tukey1985computer, wilkinson2008scagnostics} and some established time series features \citep{hyndman2021forecasting}, these indices capture variation, shape and trend, and sequential temporal behaviours of the data. The final stage utilises visual tools to identify outliers, patterns, potentially interesting features and to support interpretation of the diagnostic indices.

The \textit{wdiexplorer} package was originally designed for exploratory analysis of WDI data, one indicator at a time. We adapt some of the diagnostic indices and visualisation tools of \textit{wdiexplorer} to demonstrate how it can also be used to assess consistency, variation, and comparison between observed data and model-based estimates of any country-level panel data. Using family planning survey observations and the corresponding modelled estimates as an example, we show how different functionalities of the package are used depending on whether the focus is on exploring the data and model outputs or the residuals.

The selected diagnostic indices for exploring the family planning datasets are silhouette width and trend strength. Silhouette width \citep{rousseeuw1987silhouettes} is typically used in cluster analysis to assess the performance of groups obtained from a clustering procedure. Here, we use it to assess a pre-defined grouping of countries. It evaluates how well a country's data aligns with its pre-defined group compared to other groups. It is computed by comparing the average distance to countries within the same group against the distance to the closest neighbouring group. The silhouette width ranges from $-1$ to $+1$. A value near $+1$ indicates that the country is well aligned within its assigned group, while a value close to $-1$ suggests the country is closer to an alternative group. Silhouette width in our context quantifies the degree to which the contraceptive use pattern of each country align with its sub-region in both the survey data and the modelled output.

Trend strength quantifies the extent to which a data series follows a consistent pattern over time, which could be linear or curved, relative to random fluctuations. The trend strength is defined as the ratio of the variance of the remainder component ($R_t$) to the variance of the combined sum of the trend and remainder components ($T_t + R_t$) from a time series decomposition \citep{hyndman2021forecasting}. The value ranges from $0$ to $1$ expressed in proportions, distinguishing smooth patterns from irregular noise.

To assess model fit, we use the diagnostic indices linearity and curvature to analyse the model residuals. Linearity measures how closely a series follows a straight-line pattern over time. It assesses whether changes occur consistently without sharp curves or bends while curvature measures the extent to which a series deviates from a straight-line trend over time, capturing the presence of non-linear patterns; upward or downward bends. Linearity is quantified as $\beta_1$, the coefficient of the linear term and curvature is quantified as $\beta_2$, the coefficient of the quadratic term in a polynomial regression fitted to the trend component of the decomposed series, where the predictors are the first two orthogonal polynomials of the time index \citep{hyndman2021forecasting}. If the model perfectly captures the patterns in the data, the residuals for each country should behave randomly, and the corresponding linearity and curvature metrics should be zero or close to zero.

Other diagnostic measures of \textit{wdiexplorer} are: country average dissimilarity, within group average dissimilarity, smoothness, number of crossing points, longest flat spot and autocorrelation. These measures are less suitable for the family planning datasets for several reasons. The infrequent nature of the survey data with intervals of 3–5 years limits the comparability of temporal measures such as the number of crossing points or longest flat spot with the model-based annual estimates. Silhouette width and trend strength diagnostic measures are considered most suitable for capturing the behaviour of family planning data. We proceed to explore these two datasets, compare their diagnostic measures, and examine patterns in the residuals using linearity and curvature indices.

\section{Preliminary Data Exploration}\label{pre_data}

\subsection{Data Processing}
The family planning datasets were processed to align  with the requirements of the \textit{wdiexplorer} workflow, which expects a single observation per country-year pair. Survey data on contraceptive use and other family planning indicators were collected from multiple survey types. In 41 countries, there are multiple survey observations per country-year pairs. To address this, a yearly observation is obtained from the surveys prioritised in the following order: Demographic and Health Surveys (DHS) first, followed by Multiple Indicator Cluster Surveys (MICS), Performance Monitoring for Action (PMA) surveys, national surveys, and other sources.

We prioritise survey types in this order because DHS is generally considered the most reliable and widely comparable, serving as a global survey for health data. Analyses using FPEM have demonstrated that the random errors associated with non-DHS data are greater than those associated with DHS data \citep{alkema2013national}. MICS and PMA follow because they are also nationally representative and rigorous, while national surveys and other sources are used only when higher-priority data are unavailable, acknowledging their potentially greater measurement error and variability. Despite prioritising survey types to obtain one observation per country–year pair, 16 country-year pairs across 12 countries have two observations of the same survey type. For each country-year pair with two observations for the same survey type, the mean of the proportions was computed and used as the representative value for such pair.

For the analysis, we select the 85 focus countries of the FP2030 initiative, previously identified as the 69 priority countries of FP2020. These countries were characterised as focus countries because they represent populations with significant barriers to accessing family planning services and have the largest gaps in meeting demand for modern contraception \citep{unitednations2019}. Many are located in Western, Eastern Africa and South-Central Asia, predominantly identified by their socioeconomic status, consisting of low and lower-middle income countries. Table \ref{data_table} (see Appendix) presents the 85 focus countries along with the total number of surveys conducted and the year of the most recent survey. The dataset spans five world regions, with Africa having the largest number of countries (48) across five sub-regions, and Europe having the fewest, with only one country in Eastern Europe.

We further limited the dataset for this analysis to survey observations from 1990 through the most recent available survey year. Prior to 1990, family planning programs were primarily focused on reducing fertility and population growth, and the corresponding survey data reflect these early program goals rather than the rights-based perspective on reproductive health and well-being \citep{ramarao2015aligning}. We also limited the analysis to the population of women who are married or in a union. This group is often referred to as Married/In-Union Women of Reproductive Age \citep{cahill2020increase}.

\subsection{Visualisation of Family Planning Survey Data}
\citep{akinfenwa2024visualisation} outlined five principles for effective visualisation of panel data: (1) one panel per individual-level (country), (2) panel grouping to represent hierarchical structures, (3) the use of colour to distinguish and compare hierarchies, (4) panel ordering to facilitate comparison, and (5) consistent panel scaling for effective comparison and interpretation. These principles guided our preliminary exploration of the family planning survey data. We applied them by constructing stacked plots of modern and traditional contraceptive use, with each panel representing a country. The panels were grouped according to the UNPD sub-regional classifications and ordered by their average total contraceptive use (modern + traditional) for their most recent survey year. Countries within each sub-region were ordered by their total contraceptive use for the most recent survey year. We maintained consistent panel scaling to ensure that the differences between countries were meaningfully represented.

\begin{figure}[H]
  \centering
  \includegraphics[width=\textwidth, height=1\textheight, keepaspectratio]{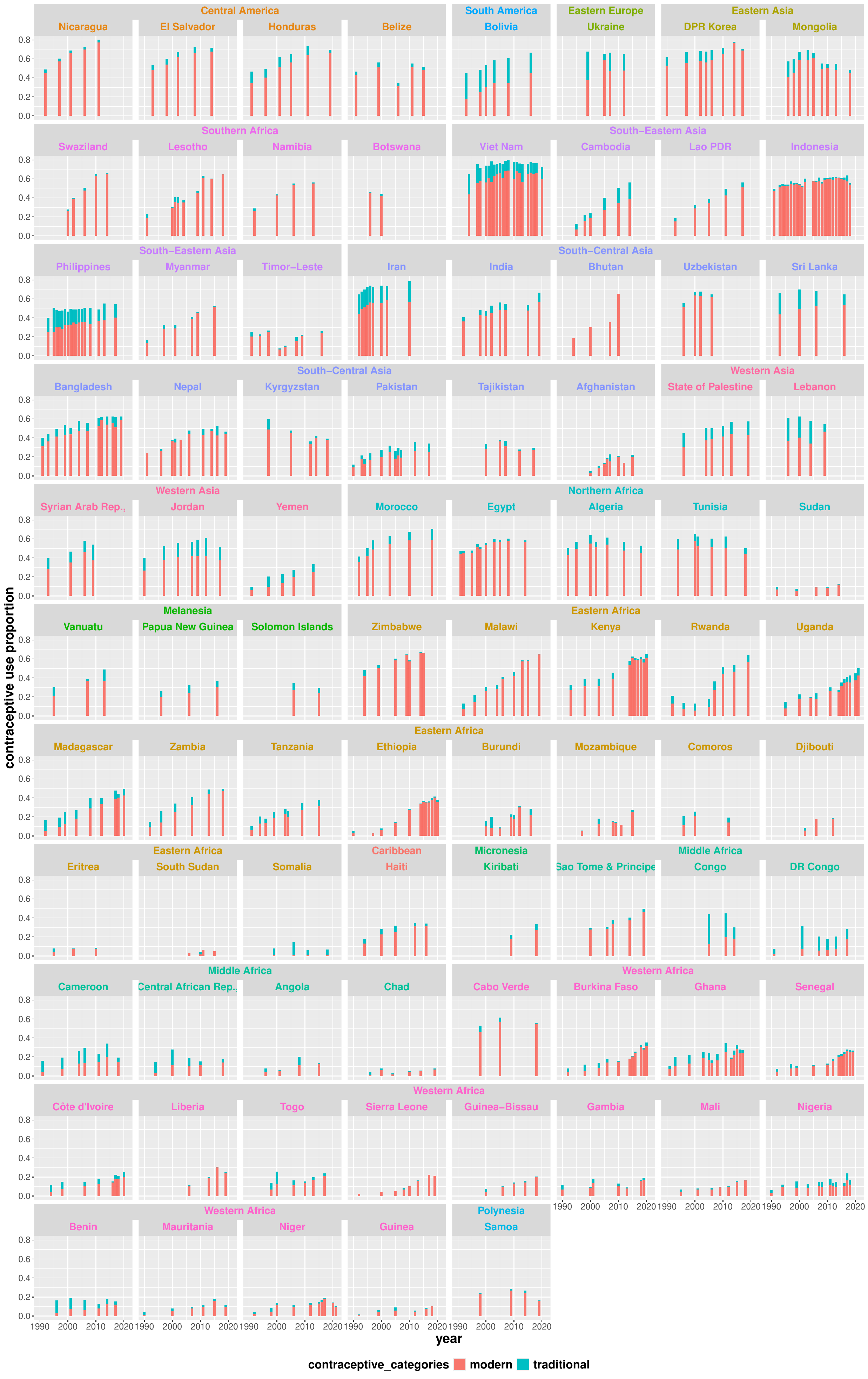}
  \caption{Stacked bar plot of contraceptive use showing both modern and traditional methods across countries. Countries are grouped by sub-region, with sub-regions ordered by their average total (modern $+$ traditional) contraceptive use in the most recent year and countries within sub-regions are also ordered according to their total contraceptive use in the most recent years. Central America sub-region has the highest (modern $+$ traditional) contraceptive use with Nicaragua having the highest total contraceptive use in its most recent survey year (2011) in Central America sub-region.}
  \label{fig:stacked-plot}
\end{figure}

Figure \ref{fig:stacked-plot} presents the proportions of modern and traditional contraceptive use among FP2030 focus countries. At the sub-regional level, Central America records the highest average contraceptive use, followed by South America and Eastern Europe. South America and Eastern Europe are each represented by a single FP2030 focus country, Bolivia and Ukraine, respectively. There are noticeable differences in total contraceptive use within and across sub-regions. In the Eastern Asia subregion there are just two FP2030 focus countries, the Democratic People's Republic of Korea and Mongolia, both of which have high total contraceptive use. In South-Eastern Asia, a few countries exhibit consistently high total contraceptive use, with Vietnam being the highest. In South-Central Asia, Iran records the highest total contraceptive use in its most recent survey year. Likewise, Morocco in Northern Africa, Zimbabwe in Eastern Africa record the highest total contraceptive use in their most recent survey year.

Figure \ref{fig:stacked-plot} also reveals variation in survey frequency across countries, as recorded in Table \ref{data_table} (see Appendix). Vietnam, Indonesia, and Philippines, countries in South-Eastern Asia, display the longest periods of annually recorded survey data. This indicates that the total number of surveys conducted in these countries between 1990 to 2021 exceeds that of any other FP2030 focus country. In contrast, some countries exhibit only short periods of annual data collection. For example, Lesotho has annual survey data between 2000 and 2002 only. Other countries, particularly countries in Central America and Melanesia sub-regions, show consistent gaps of 3 to 5 years between surveys. Beyond variation in survey frequency, Figure \ref{fig:stacked-plot} shows that some sub-regions include only a single FP2030 focus country, as the sole population in the  sub-region facing significant barriers to accessing family planning services: Bolivia in South America, Ukraine in Eastern Europe, Haiti in Caribbean, Kiribati in Micronesia and Samoa in Polynesia. 

Generally, in Figure \ref{fig:stacked-plot}, the bar plot reveals that modern contraceptive use accounts for a larger share of total contraceptive use than traditional methods. However, exceptions are observed in some countries where the proportion of traditional contraceptive use is higher. For example, the proportion of traditional contraceptive use was higher in Bolivia during its first survey year (1993), likewise all survey years in Somalia, a country in Eastern Africa. Similar pattern occurs in initial survey years across countries in Middle Africa sub-region, as well as some survey years in countries of the Western Africa sub-region. This pattern is consistent with findings reported by \cite{unitednations2019} which indicate that some countries report higher proportion of traditional contraceptive use over modern revealing barriers to accessing family planning services. 

According to the results of \cite{unitednations2019}, the median prevalence of modern contraceptive use among women of reproductive age (15–49 years) in 2019 was 44.3\% (95\% uncertainty interval: 42.1 – 47.0\%). Traditional methods accounted for approximately 4.2\% of use, calculated as the difference between overall contraceptive prevalence (48.5\%) and modern methods. \cite{marquez2018traditional} also emphasised that modern contraceptive methods are more reliable than traditional methods and are positioned higher in the hierarchy of contraceptive effectiveness. Backed by these findings, the subsequent analysis focuses on modern contraceptive use.


\section{Diagnostic Indices of Survey Data and Model-based Estimates}\label{model_eda}
FPEM relies on hierarchical Bayesian inference, which allows the model to share information across countries, sub-regions, regions and globally. The model produces smoothed estimates, and the hierarchical structure is especially useful where sparse data poses a challenge. We assess not only how well the Bayesian model captures the temporal and variational patterns observed in the survey data, but also where hierarchical pooling influences the estimates. Using diagnostic measures such as trend strength and silhouette width from the collection of diagnostic indices of the \textit{wdiexplorer} package, as described in Sub-section \ref{widexplorer}, we compare survey data and model-based estimates, and identify countries where deviations are most pronounced, highlighting contexts in which the model hierarchy has a substantive impact.

\subsection{Trend Strength Measure} \label{trend-strength}
Trend strength is one of the diagnostic indices of the \textit{wdiexplorer}, under the category of trend and shape features. It quantifies the degree to which a series follows a consistent temporal pattern, whether linear or non-linear, relative to random fluctuations. Trend strength is the proportion of variance in the data explained by the variance of the remainder component of a decomposed time series relative to the variance of the sum of the smoothed trend component and the remainder component \citep{hyndman2021forecasting}. A series is considered to have a strong trend when the variance of the remainder component is small relative to the total variance. A weakly trended series has a large remainder variance, reflecting greater irregular fluctuations \citep{akinfenwa2025wdiexplorerrpackagedesigned}.

To evaluate the strength of trends in the family planning survey data and model-based estimates, we calculate trend strength of each country in both datasets using the \textit{wdiexplorer} package. These metrics were compared by calculating the ratio of model to survey trend strength.

\begin{figure}[H]
  \centering
  \includegraphics[width=\textwidth]{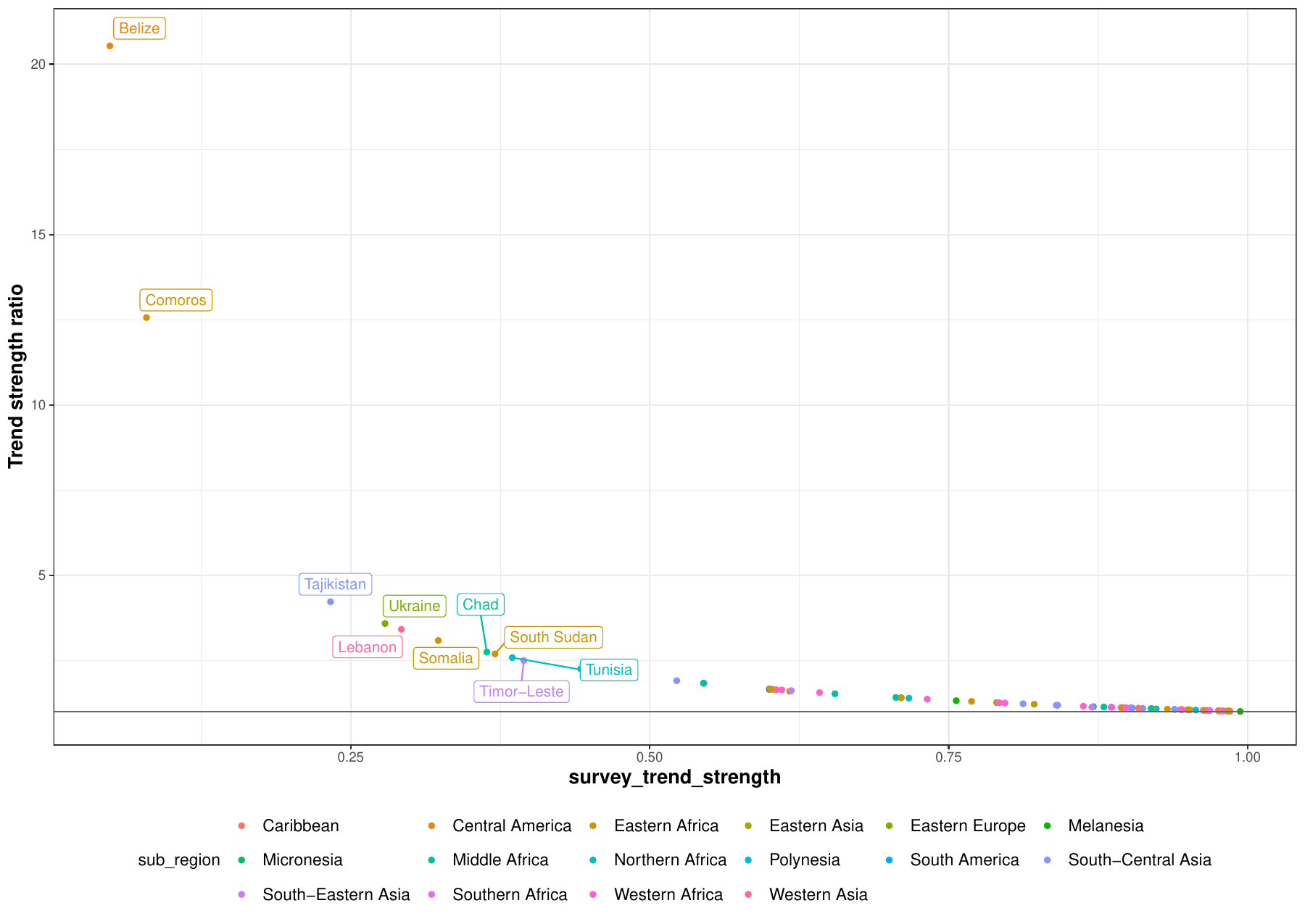}
  \caption{Comparison of trend strengths in survey data and model-based estimates across countries. Each point represents the ratio of model-based to survey trend strength plotted against the survey trend strength for a country. All countries have ratios above 1, indicating that the model-based trends are stronger than the survey trends. Labelled points correspond to countries with extreme ratios of model-based to survey trends strength. Hovering any point in the \href{https://oluwayomi-olaitan.github.io/Family-Planning-Exploratory-Analysis-Interactive-Plots/interactive-plots/trendstrength_plot.html}{interactive version} reveals the corresponding country name, its survey and model trend strength.}
  \label{fig:trendstrength-plot}
\end{figure}

We present a scatterplot of the model to survey trend strength ratio versus the survey data trend strength for each of the 85 FP2030 focus country in Figure \ref{fig:trendstrength-plot}. 

Figure \ref{fig:trendstrength-plot} shows that across all 85 focus countries, the trend strength ratio exceeds 1, indicating that the model-based trend strength is stronger than that derived directly from survey data. This pattern is consistent with the smoothing effect of FPEM. Smoothing increases apparent trend strength because it removes short-term noise and measurement error, borrows information from neighbouring time points to reinforce consistency, assumes gradual continuous change rather than abrupt shifts, and down-weights outliers; together these effects suppress variability in the raw data and make the underlying long-term pattern appear clearer, more consistent, and stronger than it does in the observed data points. Labelled points correspond to the top ten countries with the largest ratios, pointing to cases where the model estimates differ most from the observed survey trends. We further examine these countries by comparing their survey data points with their corresponding model-based trends.

\begin{figure}[H]
  \centering
  \includegraphics[width=\textwidth]{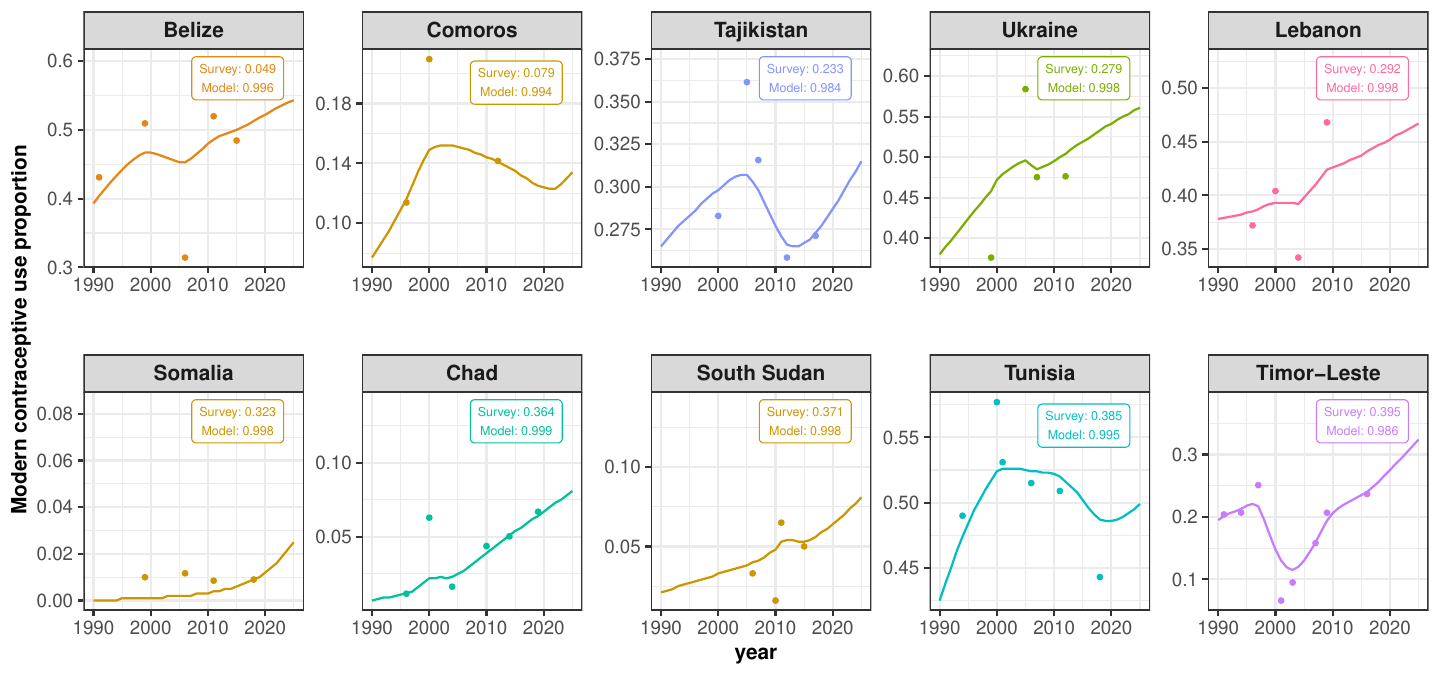}
  \caption{Survey data trajectories and model-based trends for the top 10 countries with the largest ratios of model-based to survey trend strength. Each panel presents the survey observations and corresponding model trajectory of a country, ordered by their ratios. Belize shows the highest ratio, reflecting a dip in its 2006 smoothed out by the model. Similarly, Comoros exhibits a sharp spike in 2000, with an observed proportion of modern contraceptive use far higher than other years.}
  \label{fig:trendstrength_trajectories-plot}
\end{figure}

For the 10 countries with the largest deviations, Figure \ref{fig:trendstrength_trajectories-plot} displays their survey data points alongside the corresponding model-based trends over time. Each country is represented by a separate panel, and countries are ordered in descending order of the ratio value. We ensure panel-specific scaling to accurately present the fluctuations in data points and trends for each country. Figure \ref{fig:trendstrength_trajectories-plot} illustrates how FPEM smooths survey fluctuations in the top 10 countries with the highest ratios. The smoothed curves are designed to highlight the underlying signal by filtering out some of the noise in the data. That makes the model based long-term trends look stronger and clearer than they appear in the raw observed data.

All 10 countries have trend strength values calculated from the modelled estimates that are very close to 1, as one might expect. The differing ratios reflect variation in the trend strength of the observed data. Belize has a very low trend measurement for the survey observations, and its panel shows that the proportion of modern contraceptive use decreased substantially in 2006 compared to previous years, followed by higher values in subsequent surveys (2011 and 2015). This dip in the survey data is weakly captured in the model-based trend, as the model assumes gradual, continuous change over time rather than abrupt shifts. Comoros has a similarly low trend measurement for the survey observations, but based on just three observations, where the middle one is unusually large.

Other countries with high ratios, namely Tajikistan, Ukraine, Lebanon, and South Sudan, have sparse data, which does not follow a consistent upward or downward trend. In Somalia, the fitted trend exhibits bias and lies consistently below the observed survey points. In Figure \ref{fig:stacked-plot}, Somalia exhibits the lowest proportion of modern contraceptive use across all survey years, both within its sub-region and relative to other sub-regions. Given this persistent deviation, one might expect the model-based trend to be partially pulled toward the sub-regional pattern, but the trend remains low, due to the influence of the low survey values.

The high trend scores for the models relative to the surveys across the 10 countries are attributed to FPEM including temporal smoothing and information sharing across countries within its hierarchical structure. Note also, while we rank survey types and select the data point from the highest-ranked type for each country-year during our survey data processing, FPEM uses all available  survey types to generate model estimates. Together, these modelling choices reduce the influence of survey noise, resulting in smoother trajectories and stronger apparent trends in the model-based estimates compared with the survey data.

\subsection{Silhouette Width Measure}
Silhouette width is an output of the silhouette analysis and is one of the variation features offered by \textit{wdiexplorer}. It quantifies how well a data series align with its assigned group relative to other groups \citep{rousseeuw1987silhouettes}. The silhouette value ranges from $-1$ to $+1$, where a value approaching $+1$ indicates that a country is well-aligned within its group and clearly separated from other groups, a value near $0$ suggests a country is not clearly separated from other groups, and a value near $-1$ indicates that the data trajectory of a country is closer to those of a different group.

Silhouette analyses for both the survey and model-based estimates were conducted using the complete family planning dataset for all participating countries. We calculated dissimilarities for all 185 countries, capturing the distance of each country to all others as well as its distance to countries within its sub-region. These dissimilarities formed the basis for computing the silhouette width. Although the analysis draws on the complete family planning dataset available from the UNPD website, the partition plots display only FP2030 focus countries with more than one country in their sub-region. Sub-regions with only a single FP2030 country were excluded in the partition plot.

\begin{figure}[H]
  \centering
  \includegraphics[width=\textwidth, height=0.92\textheight, keepaspectratio]{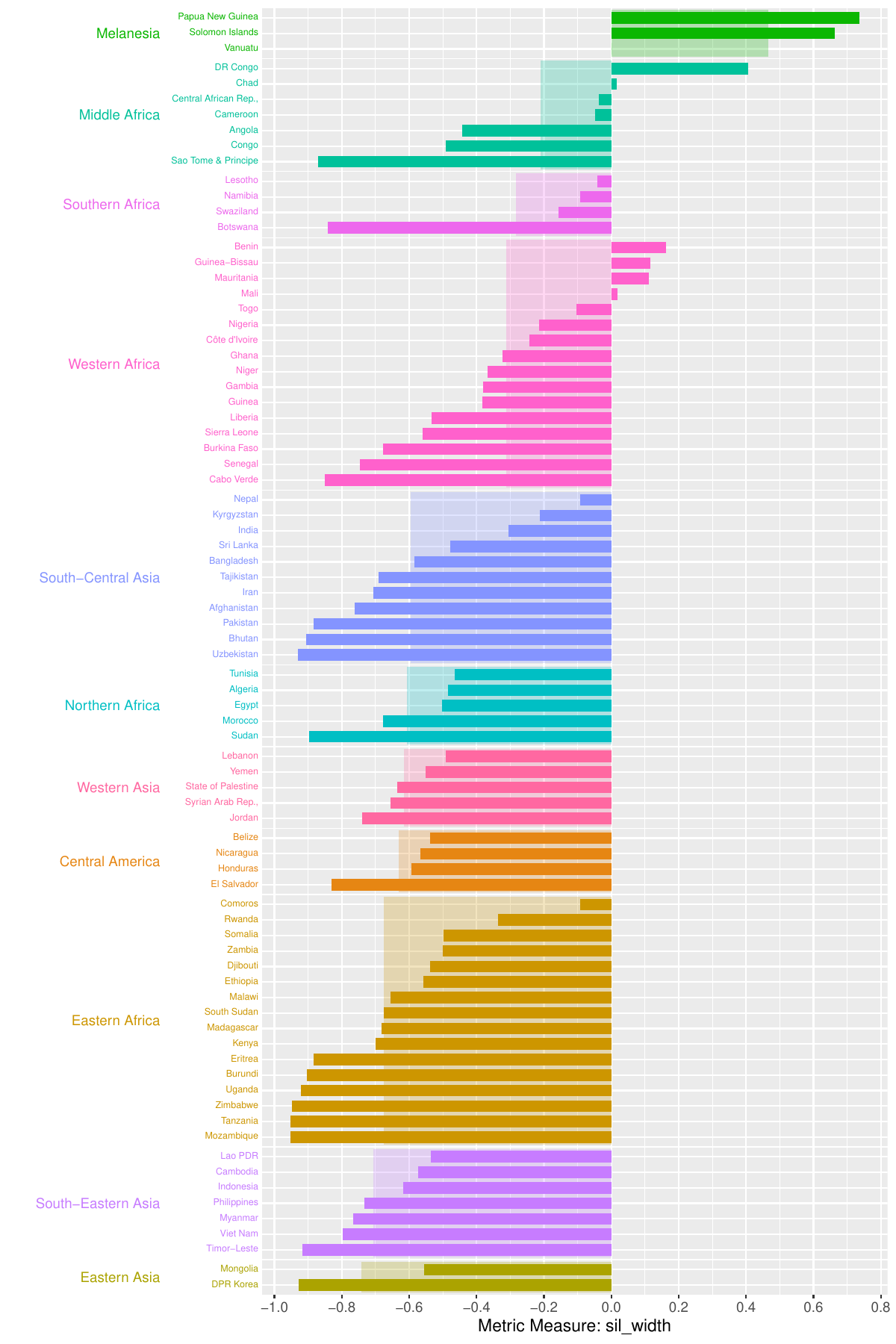}
  \caption{Silhouette widths of the survey data across the FP2030 focus countries, grouped by sub-region excluding sub-regions with only one country. Each bar represents the silhouette width of its correspondence country, while the light-shaded rectangular bars indicate the average silhouette width for each sub-region. Sub-regions are ordered by their average silhouette width, and countries within the same group are ordered accordingly with the same colour to facilitate visual distinction within groups. All sub-region except Melanesia exhibit negative group average silhouette width.}
  \label{fig:survey_silwidths-plot}
\end{figure}

\begin{figure}[H]
  \centering
  \includegraphics[width=\textwidth, height=0.92\textheight, keepaspectratio]{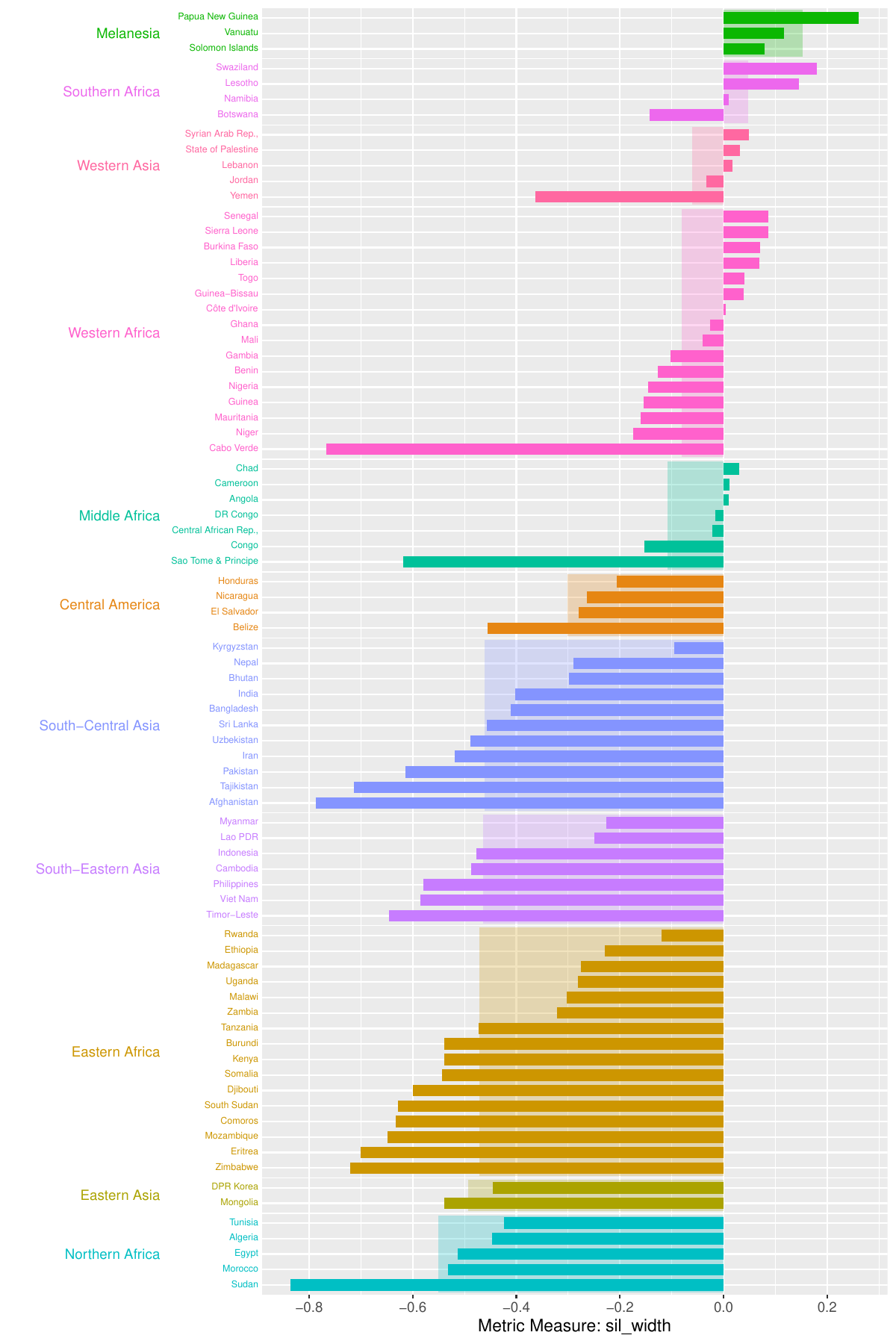}
  \caption{Silhouette widths of the model-based estimates across the 85 FP2030 focus countries, grouped by sub-region excluding sub-regions with only a single country. Sub-regions are ordered by their average silhouette width. Countries within the same sub-region share the same bar and axis text colour to facilitate visual distinction within groups. All sub-regions have negative average silhouette widths, except Melanesia and Southern Africa with positive average silhouette widths.}
  \label{fig:model_silwidths-plot}
\end{figure}

Figures \ref{fig:survey_silwidths-plot} and \ref{fig:model_silwidths-plot} are the partition plots of survey data and model-based estimates silhouette widths respectively. The partition plot is produced by one of the visualisation functions of the \textit{wdiexplorer} and displays diagnostic metrics for each country while accounting for pre-defined groupings, here sub-regions used by FPEM. Each country metric value is represented by a coloured bar over a lighter-shaded rectangular bar indicating the group average, allowing easy comparison of silhouette widths of countries within their group \citep{akinfenwa2025wdiexplorerrpackagedesigned}. 

Silhouette widths highlight how closely countries follow the patterns of their assigned groups and reveal meaningful within-group deviations. However, when a country exhibits behaviour that diverges from that of other countries in the group, average dissimilarity for the group increases, which can influence the relative alignment of countries within the group. This pattern illustrates how outlier countries can pull the group average dissimilarity, affecting the alignment of the majority of countries within the group. A silhouette analysis based on median rather than average dissimilarity may be more appropriate for our context, but we have not investigated this.

Figure \ref{fig:survey_silwidths-plot} displays the silhouette widths for the survey data of the 85 FP2030 focus countries, grouped by sub-region. Haiti, Ukraine, Kiribati, Bolivia, and Samoa are excluded from the partition plot because they are the only FP2030 country in their sub-region, leaving no other focus countries for comparison. Note, other countries do exist in their sub-regions in the complete family planning dataset used to compute the silhouette analysis, as reported in Table \ref{tab:subregion} (see Appendix). Sub-regions are arranged in descending order of their average silhouette widths, and within sub-regions, countries are ordered by decreasing silhouette widths. 

Melanesia, a sub-region in the Oceania region, is the only sub-region with a positive average silhouette width. In this sub-region, Figure \ref{fig:stacked-plot} shows that the three FP2030 focus countries consistently exhibit proportions of modern contraceptive use within a similar range across all survey years, although Vanuatu displays slightly higher levels than the other two countries. This overall similarity is reflected in their comparable and positive silhouette widths. Papua New Guinea with only three data points as reported in Table \ref{data_table} (see Appendix) has a relatively high silhouette width of $0.74$ and Solomon Islands, with two data points, has a silhouette width of $0.66$. Vanuatu has a silhouette width of $0$, because none of its surveys were conducted in the same years as those of other 5 countries in the Melanesia sub-region. (Recall that the complete dataset is used for the silhouette analysis). The group average silhouette width in this sub-region also reflects the observed low within group dissimilarity with positive average group silhouette width.

In Central America sub-region, the focus countries consistently exhibit proportions of modern contraceptive use within a similar pattern across survey years, see Figure \ref{fig:stacked-plot}. However, the silhouette widths shown in Figure \ref{fig:survey_silwidths-plot} do not reflect this observation. This is explained by differences in survey years across countries. Survey observations were conducted in different years, and in this sub-region, only a few surveys were conducted in the same year. As a result, dissimilarities for some country pairs were computed using only one or, at most, two common survey years. Silhouette widths in this sub-region are estimated solely on the years where both countries have recorded observations, which may not adequately capture the observed overall similarity.

Eastern Asia has the lowest average group silhouette width, with two of our focus countries exhibiting high negative silhouette widths. In Middle Africa, the Democratic Republic of Congo and Chad show positive silhouette widths, whereas other countries in the sub-region have negative values, with São Tomé and Príncipe exhibiting the lowest. In Southern Africa, Botswana has an extreme negative silhouette width, indicating that its trajectory of modern contraceptive use is distinct from all other countries in the sub region.

The influence of an outlier country is seen across most sub-regions. For example, in South-Eastern Asia, Timor-Leste exhibits the lowest proportion of modern contraceptive use, as shown in Figure \ref{fig:stacked-plot}. Its silhouette width, which quantifies how closely a country follows the patterns in its assigned group, indicates that Timor-Leste has the lowest value, reflecting its minimal alignment with the South-Eastern Asia sub-region (Figure \ref{fig:survey_silwidths-plot}). This extreme deviation contributes to an increased group average dissimilarity, reflecting the influence of an outlying country. A similar effect is observed in Sudan within the Northern Africa sub-region, where its contraceptive use is the least compared to the neighbouring countries, resulting in low silhouette width and higher group average dissimilarity. 

In Western Africa, Cabo Verde has the highest proportion of modern contraceptive use across its survey years relative to other countries in the sub-region. Its low silhouette width indicates that it is the least aligned country within the Western Africa sub-region, thereby contributing to a negative group average  silhouette width and thus higher overall group dissimilarity. By contrast, several countries within a similar range of proportions of modern contraceptive use in Western Africa sub-region, namely Benin, Guinea-Bissau, and Mauritania exhibit positive silhouette widths, demonstrating better alignment. 

Overall, most countries exhibit strongly negative silhouette widths. In the literal interpretation of silhouette width, these values indicate that the data series trajectories of these countries are more similar to countries outside their assigned sub-regions than to those within. Several factors may contribute to these deviations. Firstly, we observed in Figure \ref{fig:stacked-plot} that there is high variation in contraceptive use within each subregion. Furthermore, as discussed above, the presence of outlier countries can affect group alignment. In addition, family planning survey data were collected at irregular intervals, and because dissimilarities are computed only between observations from the same year, silhouette widths are estimated using dissimilarities of only those years where both countries in a pair have recorded observations. These inconsistencies could have amplified negative silhouette widths and weaken the alignment of countries within their sub-region.

Figure \ref{fig:model_silwidths-plot} is the partition plot of the model-based estimates dataset for the 85 focus countries, excluding the 5 countries that are the only country in their sub-regions. Figure \ref{fig:model_silwidths-plot} shows that all countries in the Melanesia sub-region exhibit positive silhouette widths as seen in Figure \ref{fig:survey_silwidths-plot}. Notably, Vanuatu now has a silhouette value; in the survey partition plot of Figure \ref{fig:survey_silwidths-plot}, it lacked a value because the survey years of its observations did not match the survey years of the other countries in the sub-region. With FPEM generating annual estimates, this is no longer an issue. 

Southern Africa shows a relatively small positive average silhouette width, with Botswana as the only country exhibiting a negative value, whereas in Figure \ref{fig:survey_silwidths-plot}, all countries exhibit negative silhouette widths. This reflects how the use of model-based annual estimates from FPEM, which enables consistent year-to-year comparisons across countries, can result in more robust silhouette widths that better represent alignment within this sub-region. In Western Africa, Cabo Verde stands out with an extremely low silhouette width. This pattern was also evident in Figure \ref{fig:survey_silwidths-plot}, as Cabo Verde is the only focus country in the Western Africa sub-region with a high uptake on modern contraception. Similarly, Yemen stands out among Western Asia countries with an extremely low silhouette width, though this was not apparent in Figure \ref{fig:survey_silwidths-plot}. Looking back at Figure \ref{fig:stacked-plot}, we see that Yemen has by far the lowest contraceptive use in its sub region. 

Other sub-regions including Central America, South-Central Asia, South-Eastern Asia, Eastern Africa, Eastern Asia, and Northern Africa are dominated by countries with moderate to strongly negative silhouette widths. This same pattern was observed in the survey-based silhouette width as seen in Figure \ref{fig:survey_silwidths-plot}.

In Figure \ref{fig:survey_silwidths-plot}, countries in the Western Asia sub-region show negative silhouette widths, with an average group width of $-0.61$. The model-based estimates in Figure \ref{fig:model_silwidths-plot} present a slightly different picture: while Jordan and Yemen remain negative, other countries exhibit positive (though close to zero) silhouette widths, with an average group width of $-0.06$. This comparison highlights how the annual model-based estimates of FPEM adjust the alignment of countries within the sub-region relative to the survey-based data. Building on this, we compared the silhouette widths from the model-based estimates with those from the survey data to assess how well the model reproduces the patterns observed in the survey and to highlight cases with high differences in their values.

\begin{figure}[H]
  \centering
  \includegraphics[width=\textwidth, height=0.93\textheight, keepaspectratio]{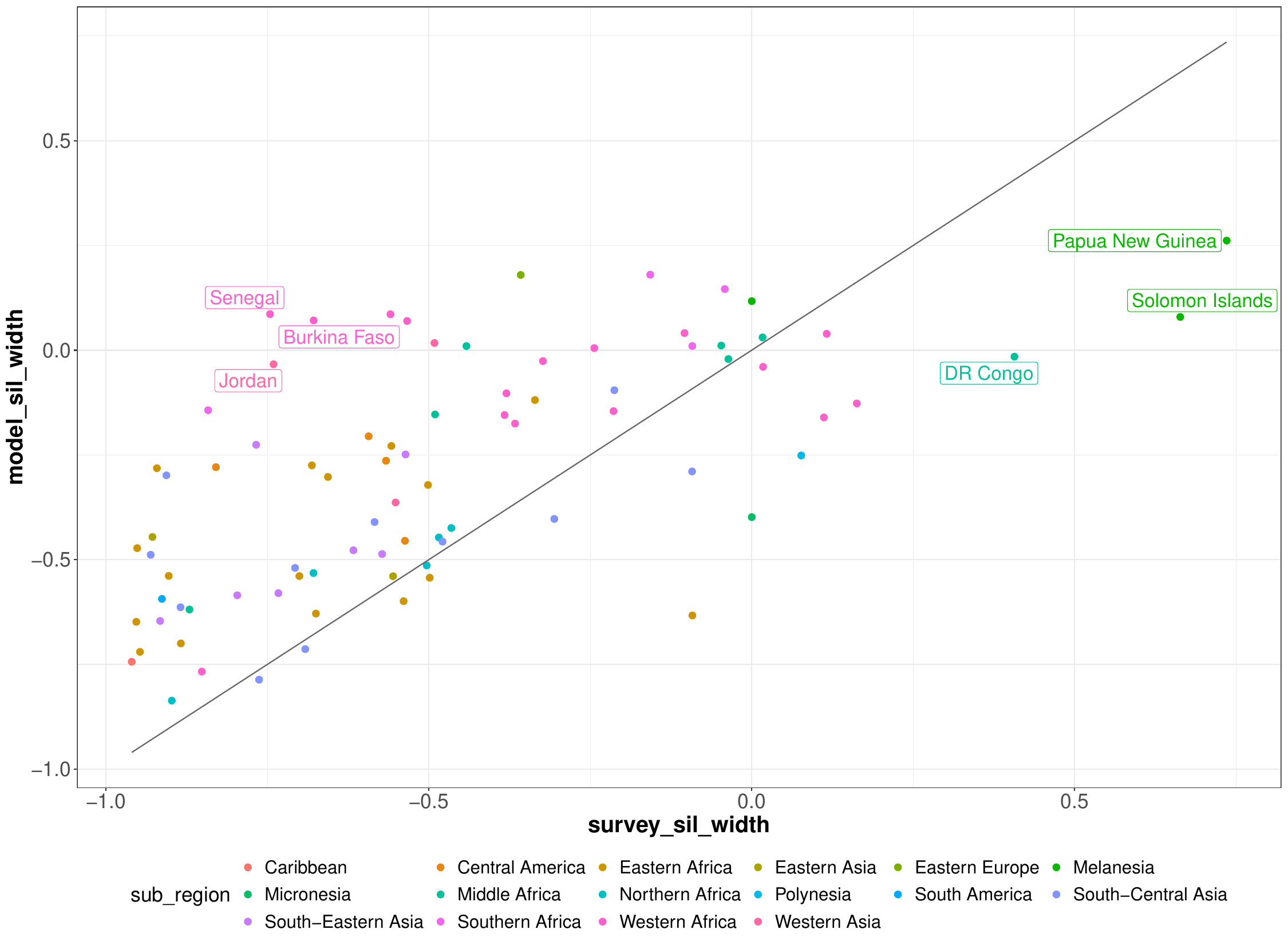}
  \caption{Comparison of silhouette widths between survey data and model-based estimates across countries. Each point represents the model-based silhouette width plotted against the survey silhouette width of a country. The diagonal line represents the identity line. Countries are coloured by sub-region to facilitate the identification of within group patterns. The labelled points correspond to the top three countries with the largest absolute differences between the two metrics and three other countries with the highest positive survey silhouette width. Countries represented by points close to the diagonal have similar silhouette widths in both the survey and model-based estimates. Hovering any point in the \href{https://oluwayomi-olaitan.github.io/Family-Planning-Exploratory-Analysis-Interactive-Plots/interactive-plots/silwidths_plot.html}{interactive version} reveals the corresponding country name, its survey and model-based silhouette widths.}
  \label{fig:silwidths-plot}
\end{figure}

\begin{figure}[H]
  \centering
  \includegraphics[width=\textwidth]{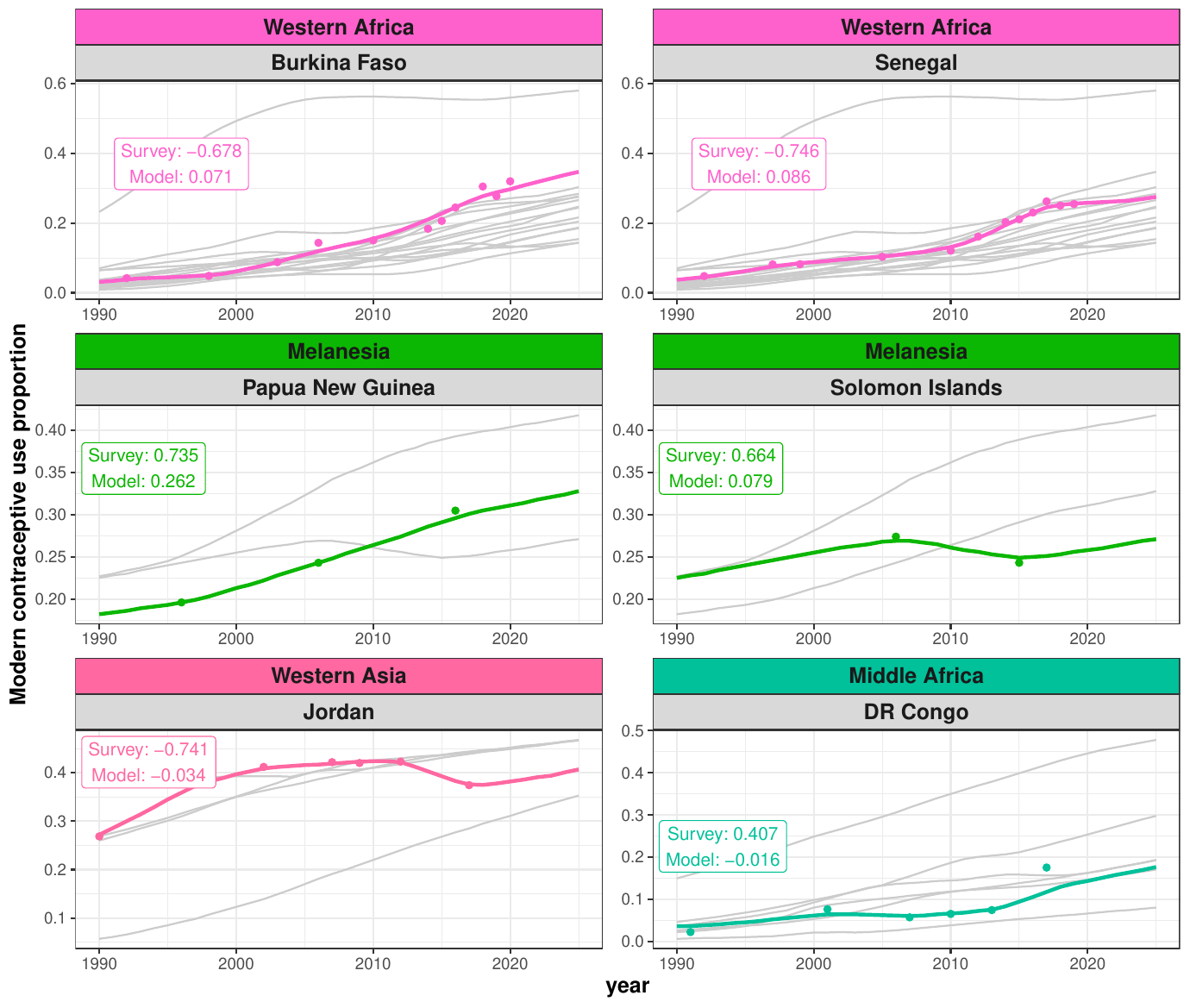}
  \caption{Survey data trajectories and model-based trends for the labelled countries with high absolute differences between model-based and survey silhouette widths in Figure \ref{fig:silwidths-plot}. The labelled countries are presented along the other countries in their sub-region to compare their data series and model trends. Sub-regions are ordered by their silhouette width. Labelled countries are uniquely highlighted by distinct colours while the model trajectory of other countries in their sub-region are presented in grey colour. Countries in Western Africa have higher model-based silhouette width, while countries in Melanesia have higher survey silhouette width. Survey and model-based silhouette widths for each labelled country are presented in their panel. Hovering any line in the \href{https://oluwayomi-olaitan.github.io/Family-Planning-Exploratory-Analysis-Interactive-Plots/interactive-plots/silwidths_trajectories_plot.html}{interactive version} reveals the corresponding country name, and its survey and model silhouette widths.}
  \label{fig:silwidths_trajectories-plot}
\end{figure}

Figure \ref{fig:silwidths-plot} presents the scatterplot of model-based silhouette widths against the survey silhouette width for each country. The labelled countries are positioned far away from the diagonal line and their observations and model-based trends are shown in more detail in Figure \ref{fig:silwidths_trajectories-plot}.

Generally, we would expect model-based silhouette values to be bigger than those calculated from the survey data. This is due to the effect of smoothing, which flattens out local discrepancies and also the influence of the Bayesian hierarchical framework which borrows strength in modelling trajectories within sub-regions. Consequently, most of the points in Figure \ref{fig:silwidths-plot} are above the diagonal line. The three labelled  countries in Figure \ref{fig:silwidths-plot} (Burkina Faso and Senegal in Western Africa, and Jordan in Western Asia) with model-based silhouette widths considerably higher than values from the survey data would consequently appear to benefit a lot from the effects of modelling, even though their model-based silhouette values are still near zero. To further examine the large differences in their silhouette widths, we compare the survey data trajectories to the corresponding model-based trends of the labelled countries.

Figure \ref{fig:silwidths_trajectories-plot} also labels three countries with unusually high survey-based silhouette width, relative to those from the model trajectories. These are the three countries with the highest silhouette width in Figure \ref{fig:survey_silwidths-plot}, and their observations and trends are shown in Figure \ref{fig:silwidths_trajectories-plot}. As noted previously and as shown in Figure \ref{fig:silwidths-plot}, the two Melanesia countries (Papua New Guinea and Solomon Islands) have just a few survey observations and so the survey-based distances and thus silhouette widths are unreliable. The third country, DR Congo based on its survey silhouette value is the most middling country in Middle Africa, which is consistent with Figure \ref{fig:stacked-plot}. The drop in silhouette value for its model trajectory is likely explained by the fact that the smooth remains flat between 2000 and 2014, unlike other Middle Africa countries.

The comparison of survey and model-based trend strengths and silhouette widths highlights the influence of the Bayesian hierarchical framework and the time series modelling in FPEM, which smooths country trajectories. The low silhouette values for the observed data reflects the fact that most of the sub-regions have quite a lot of variation across their observed data trajectories. The silhouette values calculated from the model are somewhat higher. This analysis suggests that a grouping structure other than sub-region might be more informative for family planning modelling. To further strengthen our exploratory analysis, we analysed the residuals between the model estimates and survey observations.

\section{Exploratory Residual Analysis of the Survey and Model-based Family Planning Datasets}\label{residual_analysis}

To further compare how the model-based estimates of modern contraceptive use differs from the survey data, we analyse the residuals using two selected diagnostic indices of the \textit{wdiexplorer} as discussed in Sub-section \ref{widexplorer}. We compute the linearity and curvature of the residuals for each country. If the model captures the patterns in the data trajectories, the residuals should appear as random fluctuations around the zero line, with linearity and curvature values  close to zero. When the residuals instead show some patterns, such as consistent upward or downward movements or fluctuations over time, it indicates that the fitted model has not fully captured important structure in the data.

Linearity is one of the diagnostic indices implemented in the \textit{wdiexplorer}. It measures the overall direction of a data series over time. When applied to the residuals of the family planning datasets, this measure evaluates how closely the residuals follow a straight-line pattern. The linearity value is obtained as the coefficient of the linear term $(\beta_1)$ in an orthogonal polynomial regression fitted to the smoothed trend component of the decomposed series. Positive values indicate an increasing linear pattern in the residuals, while negative values indicate a decreasing pattern. Because residuals from a well-fitting model should resemble random noise, we expect minimal to no linear structure, with linearity values near zero.

The curvature feature implemented in the \textit{wdiexplorer} measures how much a series bends away from a straight-line pattern over time, capturing non-linear movements such as gradual upward or downward curves (U- shaped and the inverted U-shaped). It is computed as the coefficient of the quadratic term of an orthogonal polynomial regression fitted to the trend component of the decomposed residuals \citep{hyndman2021forecasting}. As with linearity, residuals that behave like random noise should have curvature values close to zero.

\begin{figure}[H]
  \centering
  \includegraphics[width=\textwidth]{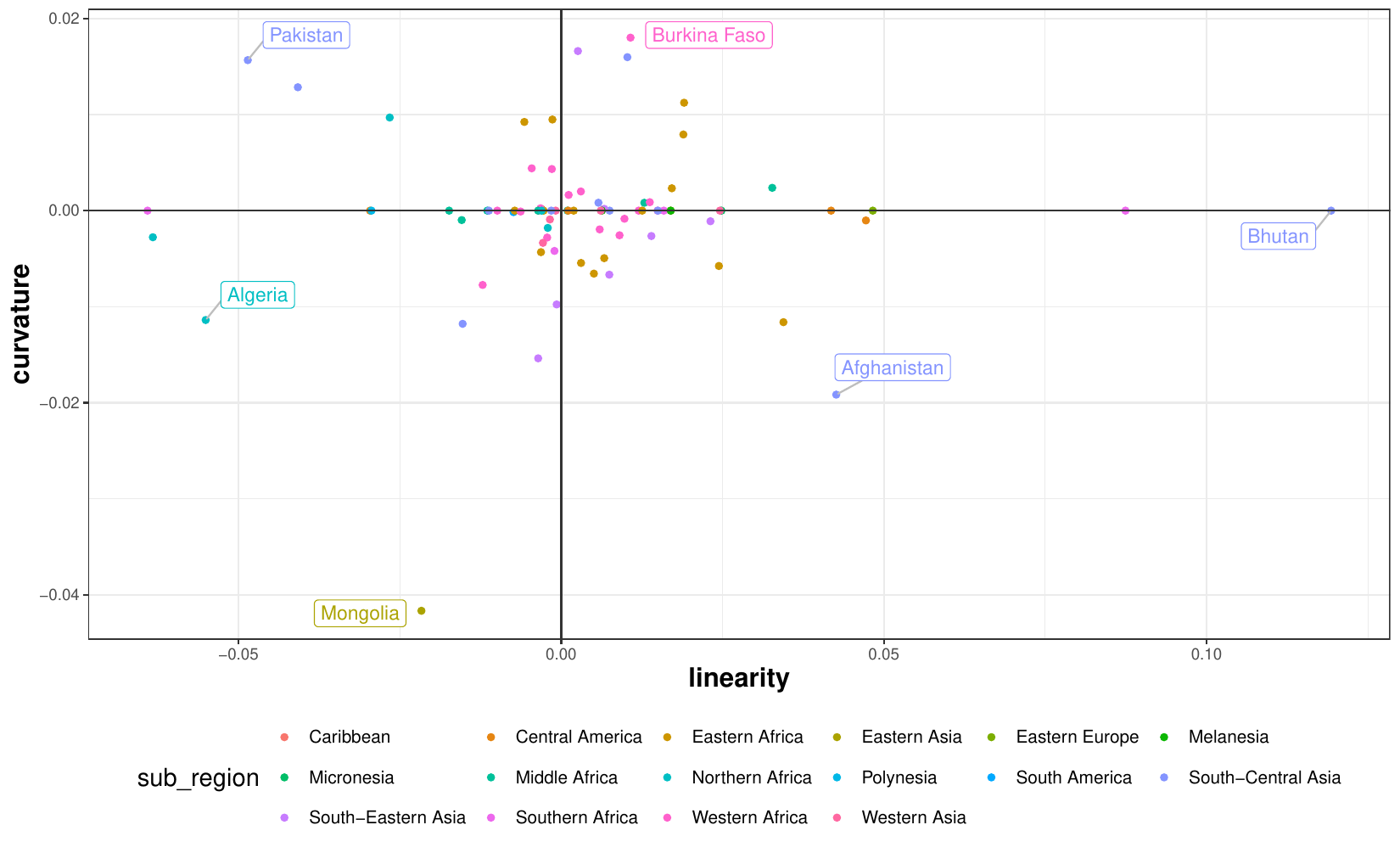}
  \caption{A scatterplot of residual curvature against linearity from the FP survey and model-based datasets, with each point representing a country and coloured by sub-region. It assesses whether linear or non-linear patterns remain in the residuals of each country. Most countries are around the vertical and horizontal lines indicating zero and near-zero curvature or linearity values. The labelled points correspond to countries with slight unexplained variation between the survey and the model-based estimates. Each panel presents the curvature and linearity values of its country. Hovering any point in the \href{https://oluwayomi-olaitan.github.io/Family-Planning-Exploratory-Analysis-Interactive-Plots/interactive-plots/curvature_linearity_plot.html}{interactive version} reveals the corresponding country name, the curvature and linearity of its residuals.}
  \label{fig:residuals-linearity-curvature-plot}  
\end{figure}

To examine the linearity and curvature of the residuals in the family planning datasets, we present a scatterplot of these two metrics in Figure \ref{fig:residuals-linearity-curvature-plot}. Most countries cluster near zero for both metrics. The labelled points indicate countries where either linearity or curvature deviates much from the expected random behaviour.

\begin{figure}[H]
  \centering
  \includegraphics[width=\textwidth]{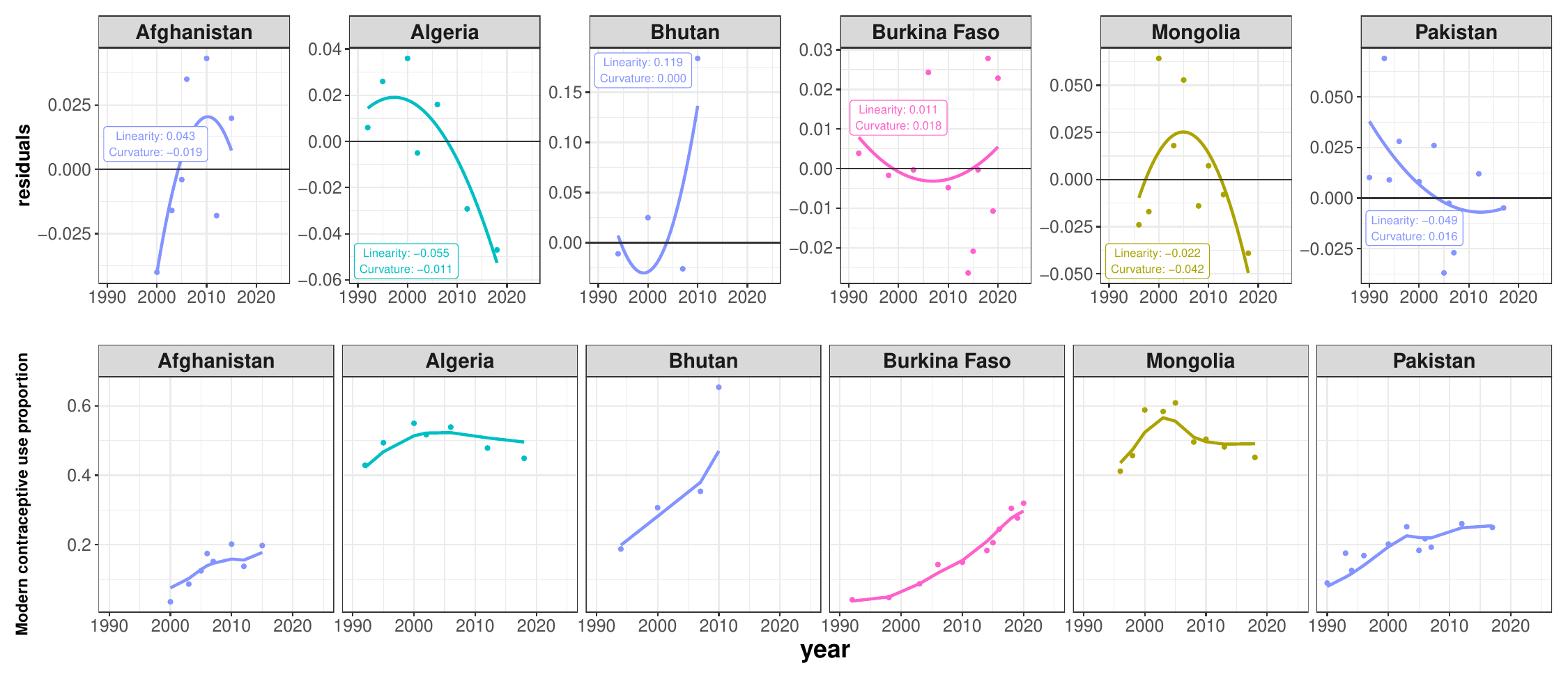}
  \caption{Labelled countries with non-zero curvature and linearity residual scores. The top panels show the residual trajectories and a superimposed quadratic fit of each country while the bottom panels display their survey data and corresponding model trends, illustrating the extent to which the model captures the patterns in each labelled country.}
  \label{fig:residuals-plot}
\end{figure}

Figure \ref{fig:residuals-linearity-curvature-plot} shows that the model performs well for most countries. The residuals cluster close to zero on both the linearity and curvature axes, indicating little evidence of systematic linear or non-linear structure in the unexplained variation. A small number of countries, which are labelled in the plot, fall somewhat farther from zero, although these deviations are not large, they show a higher absolute linearity or curvature relative to other countries. For example, Mongolia has a curvature value of $-0.042$ and a linearity value of $-0.022$, illustrating modest fluctuations in its residual pattern. In contrast, Bhutan has a curvature value of zero, while its linearity deviates noticeably from zero, suggesting a linear residual structure. Figure \ref{fig:residuals-plot} further examines these residual patterns.

Figure \ref{fig:residuals-plot} presents residual trajectories, a superimposed quadratic fit, the data series and model trends for the labelled countries from Figure \ref{fig:residuals-linearity-curvature-plot}. Burkina Faso is also included: although the residuals analysis indicates good model fit, it stood out in earlier analysis of silhouette widths for the survey and model-based trends. This inclusion illustrates a situation where strong diagnostics from residuals coexist with atypical trend-strength measures.

In Figure \ref{fig:residuals-plot}, residuals are shown in the top tow and the observed data with the model-based trends in the bottom row, with countries ordered consistently across rows. Bhutan has by far the largest linearity measure, which upon inspection is due to the substantial improvement in contraceptive use in 2010, which is not captured by the model. Upon investigation, earlier surveys in Bhutan, were based on national surveys, whereas the 2010 observation comes from a Multiple Indicator Cluster Survey. While the analysis allows for observations from different surveys and FPEM uses all surveys observations, the sudden spike was not captured by the model. In Afghanistan, the residuals exhibit a linearity of $0.043$ and a curvature of $-0.019$. The observed proportion of modern contraceptive use in 2000 is substantially lower than in other survey years, resulting in a large negative residual for this year. Similarly, in 2010, where the observed value is relatively high, the model exhibits a slight bias, which may be attributed to its hierarchical pooling effect. Together, these deviations account for the observed upward followed by downward pattern in the residuals, as captured by the linearity and curvature measures.

In Pakistan, the linearity and curvature values capture the discrepancy in the model fit over the period 2003–2012, as reflected in the residuals. Survey observations during this period show sudden increases and decreases in the proportion of modern contraceptive use, which do not follow a consistent temporal pattern. The time-series component of FPEM assumes smooth and continuous change rather than abrupt shifts; as a result, short-term fluctuations are smoothed in the model-based trends, giving rise to U-shaped residual trajectories. This behaviour is reflected in the residual curvature value of $0.016$, while the observed dip between 2003 and 2005 is indicated by the negative linearity value of $-0.049$, corresponding to a downward bias in the model during these years.

Afghanistan, Bhutan, and Pakistan, all belonging to the same sub-region, exhibit notable discrepancies that warrant further investigation of the South-Central Asia sub-region. Across these countries and the other countries examined above, residual analysis highlights discrepancies between model-based trends and survey observations. 

Somalia, a country not included in Figure \ref{fig:residuals-plot} shows a downward-biased model prediction, with the fitted trend consistently lying below the observed survey points as shown in Figure \ref{fig:trendstrength_trajectories-plot}. This downward bias is reflected in the residuals, which though positive are all small in magnitude, resulting in negligible linearity and curvature measures.

These findings establish the importance of exploring the observed data, the model-based trends, and the residuals to better understand the temporal patterns, diagnose where and investigate how well the model captures the observed patterns.

\section{Conclusion}\label{Conclusion}

This study explored survey data and FPEM-based estimates of modern contraceptive prevalence using time series diagnostic indices implemented in the \textit{wdiexplorer} package. Recognising that FPEM estimates are inherently smoothed, we assessed how well the model captures observed temporal patterns of modern contraceptive use across the 85 focus countries of the FP2030 initiative. By evaluating how well model trends reflect the underlying survey observations and the heterogeneity of family planning trajectories across countries and sub-regions, this study provides insight into the temporal dynamics of the proportions of modern contraceptive use across the 85 FP2030 focus countries and highlight where discrepancies between model-based estimates and survey observations are most pronounced, with visualisation serving as a key tool. Visualisation is central to the analysis, enabling direct comparison of relatively sparse survey observations with annual modelled trajectories and facilitating the interpretation of diagnostic measures at both country and sub-regional levels.

Across the 85 focus countries, the  selected diagnostics indices: silhouette width, trend strength, linearity and curvature indicate that FPEM generally captures long-term trends in modern contraceptive use data. Deviations from the survey data are most pronounced in countries and years exhibiting abrupt changes. For example, in Belize and Comoros, the model assumes gradual, continuous change over time rather than the sudden shifts observed in the data. In Somalia, while one would expect the model-based trend to partially align with the sub-regional patterns, it consistently falls below the observed survey points. Residual analysis highlights these discrepancies, providing a detailed view of where and how the model diverges from the observed data. Other countries with notable discrepancies, Afghanistan, Bhutan, and Pakistan, all belonging to the same sub-region, suggest that a grouping structure other than sub-region may improve the modelling of family planning trends.

This analysis is subject to limitations arising from the structure of the family planning survey data. These data are country-level survey observations obtained from multiple nationally representative household surveys, sometimes resulting in multiple observations per survey year and often temporal gaps of up to 6 years. The \textit{wdiexplorer} package is developed to explore country-level panel data with one observation per country-year pair. Also, infrequent temporal gaps make several diagnostic measures of \textit{wdiexplorer} unsuitable for capturing temporal behaviour in the proportion of modern contraceptive use. While silhouette width was considered appropriate, it poses a challenge. Dissimilarity-based measures are sensitive to temporal gaps, and could distort group-level alignment. One possible way to address this limitation is to estimate missing survey data points using linear interpolation prior to calculating dissimilarities. In addition, this analysis uses the mean dissimilarity of country pairs to quantify country-level dissimilarity; however, this measure can be sensitive to extreme distance values, an alternative approach would be the use of median instead of the mean in silhouette analysis, which may provide a more robust interpretation of group-level patterns in the data. 

In conclusion, this study demonstrates that combining time series diagnostic indices with  visualisation offers a framework for evaluating how well model-based estimates align with observations. This analysis confirms that FPEM-based estimates generally capture long-term trends in modern contraceptive use data across the 85 FP2030 focus countries, while also highlighting specific countries and periods where the model diverges from observed data, particularly in cases of abrupt changes or where sub-regional grouping may not fully capture local dynamics. By identifying these discrepancies, this approach not only provides insight into the temporal patterns of family planning trajectories but also points to opportunities for refining hierarchical modelling strategies. Despite limitations arising from sparse and unevenly spaced survey data, the integration of diagnostic measures and  visualisation enables a nuanced assessment of model performance. 

\newpage

\bibliographystyle{unsrt}
\bibliography{references}

\newpage

\section{Acknowledgments}\label{acknowledgments}

This publication has emanated from research conducted with the financial support of Taighde Éireann – Research Ireland under Grant number 18/CRT/6049. For the purpose of Open Access, the author has applied a CC BY public copyright licence to any Author Accepted Manuscript version arising from this submission.

\newpage

\section*{Appendix A: Survey Data Table}
\begin{longtable}{l c c c}
\caption{Summary of surveys for family planning focus countries} \label{data_table} \\
\hline
   \textbf{Country} & \textbf{Sub Region} & \textbf{Number of Surveys} & \textbf{Recent Survey} \\
\hline
\endfirsthead

\multicolumn{4}{c}%
{{\bfseries Table continued from previous page}} \\
\hline
 \textbf{Country} & \textbf{Sub Region} & \textbf{Number of Surveys} & \textbf{Recent Survey} \\
\hline
\endhead

\hline \multicolumn{4}{r}{{Continued on next page}} \\
\endfoot

\hline
\endlastfoot
 Afghanistan & South-Central Asia & 8 & 2018 \\
    \hline
Algeria & Northern Africa & 7 & 2018 \\
    \hline
Angola & Middle Africa & 4 & 2015 \\
    \hline
Bangladesh & South-Central Asia & 13 & 2019 \\
    \hline
Belize  & Central America  & 5 & 2015 \\
    \hline
Benin  & Western Africa  & 6  & 2017 \\
    \hline
Bhutan & South-Central Asia & 4  & 2010 \\
    \hline
Bolivia & South America & 6 & 2016 \\
    \hline
Botswana & Southern Africa & 2 & 2017 \\
    \hline
Burkina Faso & Western Africa & 11 & 2020 \\
    \hline
Burundi & Eastern Africa & 7 & 2016 \\
    \hline
Cabo Verde & Western Africa & 3 & 2018 \\
    \hline
Cambodia & South-Eastern Asia & 6 & 2014 \\
    \hline
Cameroon & Middle Africa & 7 & 2018 \\
    \hline
Central African Republic & Middle Africa & 5 & 2018 \\
    \hline
Chad & Middle Africa & 6 & 2019 \\
    \hline
Comoros & Eastern Africa & 3 & 2012 \\
    \hline
Congo & Middle Africa & 3 & 2014 \\
    \hline
Côte d’Ivoire  & Western Africa & 8 & 2020 \\
    \hline
DPR Korea & Eastern Asia & 8 & 2017 \\
    \hline
DR Congo & Middle Africa & 6 & 2017 \\
    \hline
Djibouti & Eastern Africa & 3 & 2012 \\
    \hline
Egypt & Northern Africa & 10 & 2014 \\
    \hline
El Salvador & Central America & 5 & 2014 \\
    \hline
Eritrea & Eastern Africa & 3 & 2010 \\
    \hline
Ethiopia & Eastern Africa & 12 & 2020 \\
    \hline
Gambia & Western Africa & 7 & 2019 \\
    \hline
Ghana & Western Africa & 13 & 2017 \\
    \hline
Guinea & Western Africa & 6 & 2018 \\
    \hline
Guinea-Bissau & Western Africa & 5 & 2018 \\
    \hline
Haiti & Caribbean & 5 & 2016 \\
    \hline
Honduras & Central America & 6 & 2019 \\
    \hline
India & South-Central Asia & 8 & 2019 \\
    \hline
Indonesia & South-Eastern Asia & 25 & 2018 \\
    \hline
Iran & South-Central Asia & 9 & 2010 \\
    \hline
Jordan & Western Asia & 7 & 2017 \\
    \hline
Kenya & Eastern Africa & 11 & 2020 \\
    \hline
Kiribati & Micronesia & 2 & 2018 \\
    \hline
Kyrgyzstan & South-Central Asia & 5 & 2018 \\
    \hline
Lao PDR & South-Eastern Asia & 5 & 2017 \\
    \hline
Lebanon & Western Asia & 4 & 2009 \\
    \hline
Lesotho & Southern Africa & 9 & 2018 \\
    \hline
Liberia & Western Africa & 4 & 2019 \\
    \hline
Madagascar & Eastern Africa & 9 & 2020 \\
    \hline
Malawi & Eastern Africa & 9 & 2019 \\
    \hline
Mali & Western Africa & 7 & 2018 \\
    \hline
Mauritania & Western Africa & 6 & 2019 \\
    \hline
Mongolia & Eastern Asia & 9 & 2018 \\
    \hline
Morocco & Northern Africa & 6 & 2018 \\
    \hline
Mozambique & Eastern Africa & 6 & 2015 \\
    \hline
Myanmar & South-Eastern Asia & 6 & 2015 \\
    \hline
Namibia & Southern Africa & 4 & 2013 \\
    \hline
Nepal & South-Central Asia & 10 & 2019 \\
    \hline
Nicaragua & Central America & 5 & 2011 \\
    \hline
Niger & Western Africa & 10 & 2021 \\
    \hline
Nigeria & Western Africa & 12 & 2018 \\
    \hline
Pakistan & South-Central Asia & 11 & 2018 \\
    \hline
Papua New Guinea & Melanesia & 3 & 2016 \\
    \hline
Philippines & South-Eastern Asia & 17 & 2017 \\
    \hline
Rwanda & Eastern Africa & 8 & 2019 \\
    \hline
Samoa & Polynesia & 4 & 2019 \\
    \hline
Sao Tome and Principe & Middle Africa & 5 & 2019 \\
    \hline
Senegal & Western Africa & 12 & 2019 \\
    \hline
Sierra Leone & Western Africa & 8 & 2019 \\
    \hline
Solomon Islands & Melanesia & 2 & 2015 \\
    \hline
Somalia & Eastern Africa & 4 & 2018 \\
    \hline
South Sudan & Eastern Africa & 4 & 2015 \\
    \hline
Sri Lanka & South-Central Asia & 4 & 2016 \\
    \hline
State of Palestine & Western Asia & 6 & 2019 \\
    \hline
Sudan & Northern Africa & 5 & 2014 \\
    \hline
Swaziland & Southern Africa & 5 & 2014 \\
    \hline
Syrian Arab Republic & Western Asia & 4 & 2009 \\
    \hline
Tajikistan & South-Central Asia & 5 & 2017 \\
    \hline
Tanzania & Eastern Africa & 8 & 2015 \\
    \hline
Timor-Leste & South-Eastern Asia & 8 & 2016 \\
    \hline
Togo & Western Africa & 6 & 2017 \\
    \hline
Tunisia & Northern Africa & 6 & 2018 \\
    \hline
Uganda & Eastern Africa & 12 & 2021 \\
    \hline
Ukraine & Eastern Europe & 4 & 2012 \\
    \hline
Uzbekistan & South-Central Asia & 4 & 2006 \\
    \hline
Vanuatu & Melanesia & 3 & 2013 \\
    \hline
Vietnam & South-Eastern Asia & 21 & 2020 \\
    \hline
Yemen & Western Asia & 5 & 2013 \\
    \hline
Zambia & Eastern Africa & 6 & 2018 \\
    \hline
Zimbabwe & Eastern Africa & 7 & 2015
\end{longtable}

\begin{table}[H]
    \centering
    \caption{A tabular representation of the total number of countries in each sub-region, grouped by FP2030 focus country status.}
    \begin{tabular}{c|c|c}
Sub\_region & Number of FP2030 countries & Others \\
\hline
    Eastern Africa & 16 & 2 \\
    \hline
Western Africa & 16 & 0 \\
\hline
South-Central Asia & 11 & 3 \\
\hline
Middle Africa  & 7 & 2 \\
\hline
South-Eastern Asia & 7 & 3 \\
\hline
Northern Africa & 5 & 6 \\
\hline
Western Asia & 5 & 11 \\
\hline
Central America & 4 & 4 \\
\hline
Southern Africa & 4 & 1 \\
\hline
Melanesia & 3 & 3 \\
\hline
Eastern Asia & 2 & 5 \\
\hline
Caribbean & 1 & 11 \\
\hline
Eastern Europe & 1 & 9 \\
\hline
Micronesia & 1 & 4 \\
\hline
Polynesia & 1 & 3 \\
\hline
South America & 1 & 11 \\
\hline
    \end{tabular}
    \label{tab:subregion}
\end{table}

\end{document}